\input epsf

\documentclass[draft]{aa}

\def\ltsim{\mathrel{\hbox{\rlap{\hbox{\lower3pt\hbox{$\sim$}}}\raise2pt\hbox{$<$}}}}

\begin{document}

\thesaurus{07(08.05.2, 08.18.1)} 
 
\title
{A constraint on the angular momentum evolution of Be stars}

\author{I.~A.~Steele}                   
                                                            
\institute{Astrophysics Research Institute, Liverpool John Moores University, 
Egerton Wharf, Birkenhead, L41 1LD, United Kingdom
}

\offprints{I. A. Steele}
\mail{ias@astro.livjm.ac.uk}

\date{Received 29 September 1998 / Accepted 30 October 1998}

\titlerunning{The angular momentum evolution of Be stars}
\authorrunning{Steele}
\maketitle 

\begin{abstract} 

The Be star sample of Steele et al. (\cite{s98}) has a 
distribution of $v \sin i$'s that is
luminosity dependant, with giants having lower projected rotational
velocities than dwarfs.  We show that this effect can be understood
simply in terms of angular momentum conservation during the evolution
from dwarf to giant.  Any decretion disk
or other angular momentum losing mechanism for such objects must
cause a loss of no more than $15$\% of the stellar angular momentum
over the Be phase lifetime.

\end{abstract}


\section{Introduction}
 
The importance of understanding the relationships between
rotational velocity, evolutionary status and stellar temperature in
causing the
the Be phenomenon has long been understood (Slettebak \cite{s82}). 
Recently Zorec \& Briot
(\cite{zb97}) presented evidence based on a careful evaluation of
the statistics of B and Be stars in the Bright Star Catalogue (Hoffleit \&
Jaschek \cite{bsc}) that after correction for various selection
effects there were no apparent differences in the
spectral type distribution and frequency of Be stars with respect 
to luminosity class.  In addition they showed the shape of
the $v \sin i$ distribution with spectral type was not luminosity
dependant, implying little or uniform angular momentum loss
from such objects over their lifetimes.  Here we extend that work by
using the sample of Steele et al. (\cite{s98}) to 
quantify
the evolution of angular momentum between the dwarf and giant
stages of Be stars.
We show that either conservation of angular momentum or an accumulated
loss
of up to (but no more than) 15\% of the stellar angular momentum
is allowed during the main sequence + giant lifetime
of Be stars.

\section{Distribution functions}

\subsection{Description of the sample}

In Steele et al. (\cite{s98}) we presented optical spectra of a
sample of 58 Be stars.  The sample contains objects from
O9 to B8.5 and of luminosity classes III (giants) to V (dwarfs), 
as well as three
shell stars (which we neglect for the purposes of this paper as
they have uncertain luminosity classes).  A spectral type and value
of $v \sin i$ was derived for each object in the sample.
The sample is termed a ``representative'' sample, in that it
was selected in an attempt to contain several objects that were 
typical of each spectral
and luminosity class in the above range.  It therefore does {\em not}
reflect the spectral and luminosity class space distribution of Be stars,
but only the average properties of each subclass in temperature and
luminosity.
The distributions of $v \sin i$ within each temperature and
luminosity class were carefully investigated and the conclusion
drawn that there were no significant selection effects biasing the
average properties of the objects.
However it was apparent that for all
spectral sub-types the giants had significantly lower values
of $v \sin i$ than the dwarfs. 

\def\epsfsize#1#2{0.8#1}
\begin{figure}
\setlength{\unitlength}{1.0in}
\centering
\begin{picture}(3.0,6.1)(0,0)
\put(-0.0,-0.0){\epsfbox[0 0 2 2]{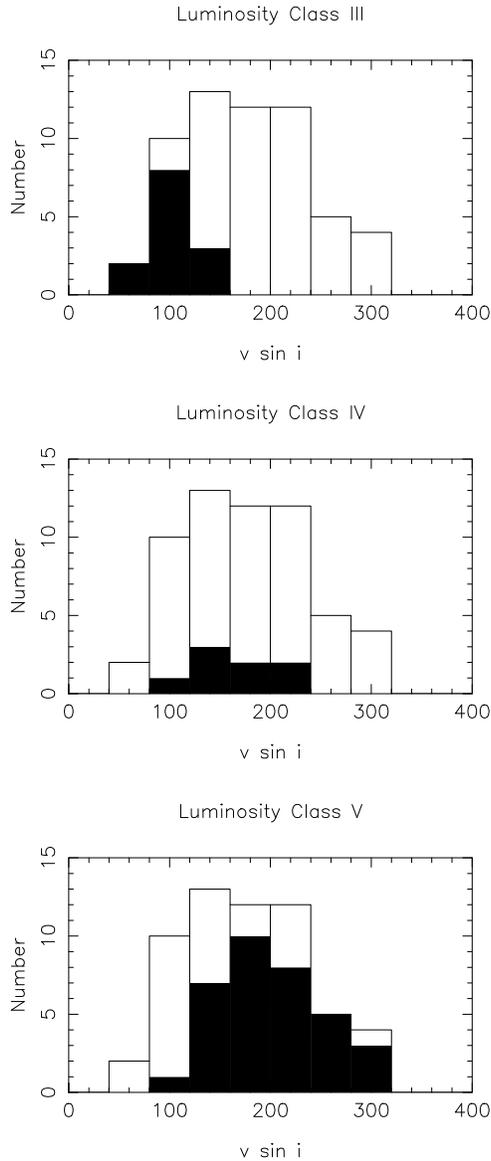}}
\end{picture}
\caption{$v \sin i$ distribution for the three luminosity classes (solid areas)
compared with the all luminosity class distribution (hollow).  A KS test shows
that the probability of the giant and dwarf distributions being
drawn from the same population is $<10^{-6}$}

\end{figure}

\subsection{$v \sin i$ and $\omega \sin i$ distributions}

In Fig. 1 we plot the binned distribution of $v \sin i$ values for the sample
for luminosity classes III, IV and V.   The data have been binned into
bins of width 80 km/s, chosen to be considerably larger than the mean error 
on any one $v \sin i$ measurement, which is $\sim 20$ km/s. It is immediately
apparent that the distributions are different, with the giants having
considerably lower $v \sin i$ than the dwarfs.  A simple explanation of this
would be that the critical velocity for giants is lower than that
for dwarfs, so that the $v \sin i$ may be lower but still give a
sufficiently high $\omega \sin i (=v \sin i / v_{\rm crit})$ to
cause a disk to form.  To investigate this we plot in Fig. 2 the
$\omega \sin i$ distributions for our sample.  We calculated $v_{\rm crit}$
according to the prescription given by Porter (\cite{p96}):
\begin{equation}
v_{\rm crit} = \sqrt{0.67 \times GM/R}
\end{equation}
where $R$ is the polar radius.  Values of $M$ and $R$ were obtained from
Schmidt-Kaler (\cite{sk82}), with interpolation between luminosity classes and
spectral sub-types where necessary. 
From Fig. 2 it is apparent that this simple explanation of
the necessity of a certain fractional velocity to give the Be 
phenomenon is insufficient to explain the discrepancy between the
giants and dwarfs.  

\def\epsfsize#1#2{0.8#1}
\begin{figure}
\setlength{\unitlength}{1.0in}
\centering
\begin{picture}(3.0,6.1)(0,0)
\put(-0.0,-0.0){\epsfbox[0 0 2 2]{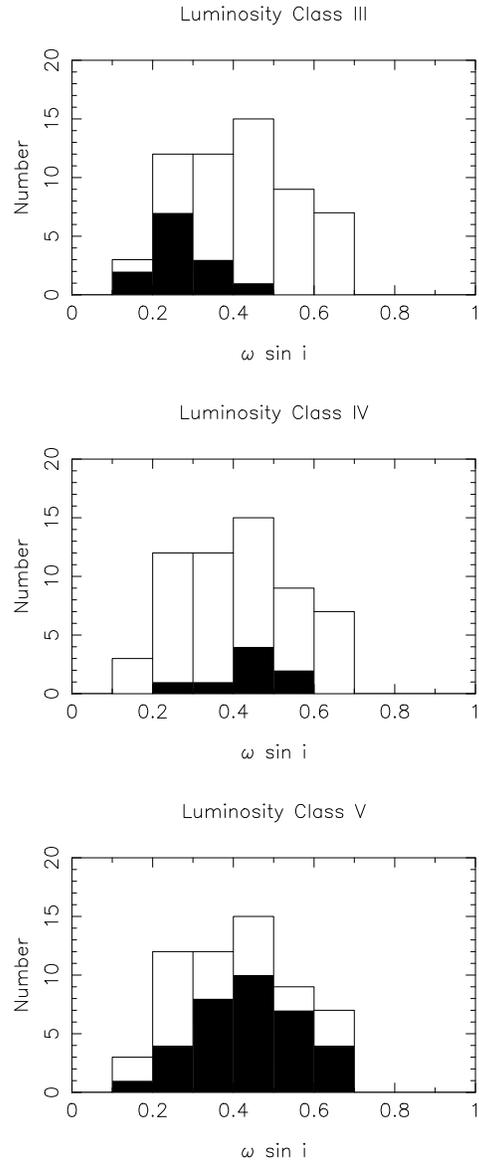}}
\end{picture}
\caption{$\omega \sin i$ distribution for 
the three luminosity classes (solid areas)
compared with the all luminosity class distribution (hollow).  A KS test shows
that the probability of the giant and dwarf distributions being
drawn from the same population is $<10^{-4}$}

\end{figure}

\subsection{Angular momentum distribution}
 
We now consider the rotational
velocity changes that result from angular momentum
conservation during the evolution from dwarfs to giants.
Assuming that the mass of a given star is fixed during this evolution
and that angular momentum is conserved, then velocity $v$ will simply
be inversely proportional to radius $R$.  The quantity we therefore
consider is $v \sin i \times R/R_{g}$ where $R/R_{g}$ is the fractional
radius for luminosity class compared to the corresponding 
giant radius.
From Schmidt-Kaler (\cite{sk82})
it is apparent that for dwarfs in 
the range O9 to B9
for a constant mass ({\em not} spectral type) 
the relationship $R/R_{g} = 1/1.8$ 
holds to within $\sim 5$ per-cent.  Similarly for the subdwarfs we adopt 
$R/R_g=1/1.4 $.  The ratio is of course unity for the giants.

In Fig. 3 we
plot the distributions of 
$v \sin i \times R/R_{g}$ for all three luminosity classes.
The similarity of the three distributions is striking.
In order to confirm their similarity we carried out a 
Kolmogorov-Smirnov (KS) test 
between the unbinned values of $v \sin i \times R/R_{g}$ for the
giants and the dwarfs.  As noted in the captions of Figs. 1 and
2 the test was also carried out
on the $v \sin i$ and $\omega \sin i$ datasets to demonstrate that
they were significantly different.  
For $v \sin i \times R/R_{g}$ the probability that the giant and dwarf
distributions are drawn from the same population is 0.83, confirming our
opinion of the similarity of the samples, and demonstrating that
the similarity was not an effect of our binning the data.  It is therefore
apparent that conservation of angular momentum over the Be lifetime of
the object is entirely consistent with the observed angular momentum
distributions.

\def\epsfsize#1#2{0.8#1}
\begin{figure}
\setlength{\unitlength}{1.0in}
\centering
\begin{picture}(3.0,6.1)(0,0)
\put(-0.0,-0.0){\epsfbox[0 0 2 2]{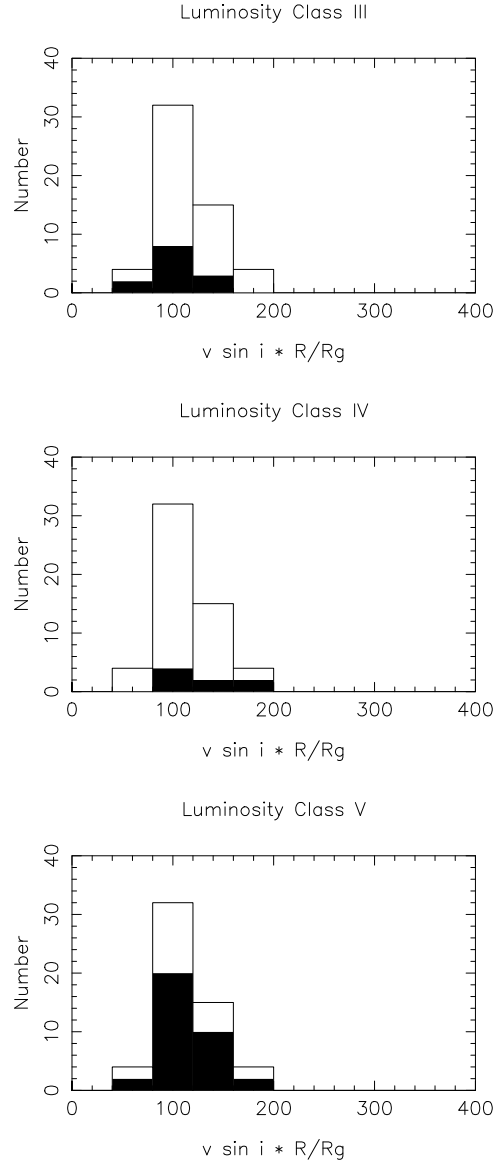}}
\end{picture}
\caption{$v \sin i \times R/R_g$ (a measure of relative 
angular momentum) distribution 
for the three luminosity classes (solid areas)
compared with the all luminosity class distribution (hollow).  A KS test shows
that the probability of the giant and dwarf distributions being
drawn from the same population is $0.83$}
\end{figure}

\section{Angular momentum evolution}

In Sect. 2.3 we demonstrated that angular momentum conservation was
consistent with the observed values of $v \sin i \times R/R_g$.
However it may be that a certain fraction of angular momentum may
be lost from the stars and the two distributions still remain consistent.
To investigate this we simulated the effect of changing the system angular
momentum of the giants by factors of between 0.01 and 2.0 in increments of
0.01 and redoing the KS test.  The resulting distribution of probabilities
is shown in Fig. 4.

\def\epsfsize#1#2{0.8#1}
\begin{figure}
\setlength{\unitlength}{1.0in}
\centering
\begin{picture}(3.0,3.3)(0,0)
\put(-0.0,-0.5){\epsfbox[0 0 2 2]{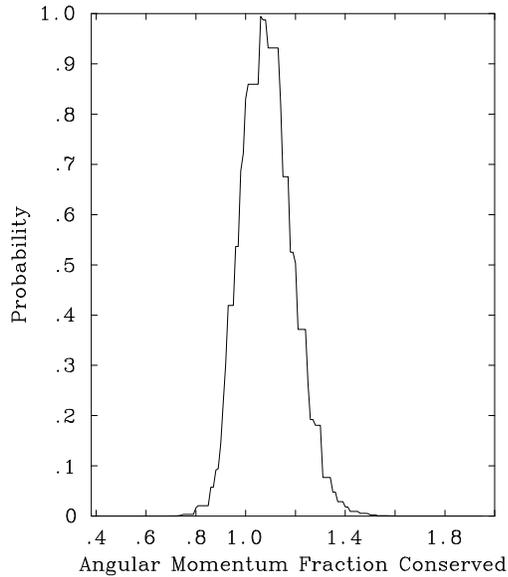}}
\end{picture}
\caption{Variation in KS test null hypothesis probability between
giant and dwarf angular momentum distributions
versus amount of stellar angular momentum conserved.}
\end{figure}

From Fig. 4 it is apparent that a probability of greater than $\sim 5$\%
of the two distributions being consistent is obtained for fractional
changes of angular momentum during the main sequence + giant
lifetime of the star of between $\sim 0.85$ and $\sim 1.3$.  
Neglecting the upper value as unphysical
we therefore conclude that any method of losing
angular momentum that purports to explain the Be phenomenon must 
cause a loss of less than $\sim 15$ per cent of the stellar angular
momentum over the main sequence + giant lifetimes of the star.
From the analysis presented by Porter (\cite{p98}) 
of the spin down of Be stars due to
angular momentum transfer to the disk (i.e. a decretion disk - 
e.g. Lee et al. 1991) 
this implies that
(assuming the Be phenomenon is present for most of the main sequence
life of the star)
the disks around Be stars are in his terminology ``weak'' to ``medium''.  This
means that for a typical disk opening angle of 15$^\circ$ and a density
of 2$\times10^{-11}$ g cm$^{-2}$ (Waters \cite{w86}), 
the initial outflow velocity
must be less than 0.01 km/s.  For a decretion disk this implies
the viscosity parameter $\alpha < 0.01$ (Porter \cite{p98}).  

An alternative
of explanation is a much ``stronger'' disk that is only present for short
periods during the life of the star.  For example if the disk were only
present for 10\% of the main-sequence lifetime, then we
derive $\alpha \sim 0.1$.

\section{Conclusions}

By using the distribution of $v \sin i$ values for giants and dwarfs in the
Be star sample of Steele  et al. (\cite{s98}) 
we have shown that any angular momentum
loss in the system that would spin down the Be stars must cause 
the loss of no more than 15\% of the stellar angular momentum.  This
implies that either the Be phenomenon is only a short phase in the
life of such objects, or that any decretion disk in the system must
have a low outflow velocity ($<0.01$ km/s) and hence a low viscosity 
($\alpha <0.01)$.

\begin{acknowledgements}

Thanks to Dr. John Porter for both his advice and his 
careful reading of the first draft of this paper.  

\end{acknowledgements}

\end{document}